\documentstyle[12pt,psfig]{article}
\def\reference{\parskip 0pt\par\noindent\hangindent 0.5 truecm}

\def\sun{\hbox{$\odot$}}
\begin{document}
%
%
\title{Location of Weak CSS Sources on the Evolutionary Path of Radio-Loud AGN}
%


\author{Andrzej Marecki$^{1}$,
 Ralph E. Spencer$^{2}$ and
 Magdalena Kunert$^{1}$
} 

\date{}
\maketitle

{\center
$^1$ Toru\'n Centre for Astronomy, Nicholas Copernicus University, Toru\'n, Poland\\[3mm]
$^2$ Jodrell Bank Observatory, University of Manchester, UK\\[3mm]
}

%
\begin{abstract}

According to the currently accepted paradigm compact steep spectrum (CSS) sources are precursors of larger/older
objects and gigahertz peaked spectrum (GPS) sources may be earlier stages of CSS sources. In this paper we
confront this paradigm with the outcome of recent observations of CSS sources that are significantly weaker than
those known before. In particular we claim that not all CSS sources must end up as large scale objects; if the
activity phase of an AGN's central engine is shorter than the lifetime of a large scale radio source (up to $\sim
10^8$~years) the radio source associated with such an AGN decays earlier, e.g.\ at the CSS stage. We point out
that a theory of thermal--viscous instabilities in the accretion disks of AGN may explain many features of radio
sources at all stages of their evolution.

\end{abstract}


\bigskip

\section{Introduction}

After more than 20 years of investigation of GPS/CSS sources (and three workshops devoted to them!) the following
paradigm appears to be stable: they are young and they grow --- GPS sources evolve towards CSS sources and these
in turn towards large scale objects (LSOs). The `frustration' scenario, although plausible, is not the principal
explanation of why CSS sources are small. This picture seems to be particularly clear if we deal with galaxies.

It looks as though the following issues still remain open:

\begin{enumerate}

\item What type of LSOs do CSS sources actually precede: only FR\,II, or either FR\,I
or FR\,II?

\item Do all CSS sources reach the LSO stage? Do all GPS sources reach the CSS stage?

\item What causes activity re-occurrences and how common a phenomenon is this?

\end{enumerate}

Based on recent observations of new samples of significantly weaker sources and the theory of thermal--viscous
instabilities in accretion disks (Hatziminaoglou, Siemiginowska, \&~Elvis 2001, hereafter HSE01, and references
therein) we present here a set of consistent answers to these questions.

\section{Only FR\,II or either FR\,I or FR\,II?}

If a CSS source is not beamed toward the observer we perceive it as a medium-sized symmetric object (MSO). It is a
well known fact based on observations of the strongest CSS sources/MSOs that morphologically MSOs resemble
Fanaroff--Riley type II objects (Fanaroff \& Riley 1974), i.e. they are double-lobed and edge brightened because
of the presence of hot spots. In other words CSS sources are mini-FR\,IIs and, surprisingly, hardly any
mini-FR\,Is are observed. Akujor \& Garrington (1995) presented a luminosity vs linear size diagram (Figure 1
therein) for the revised 3C sample which shows very clearly that the luminosities of CSS sources from the 3C
sample are of the same order of magnitude as FR\,IIs which, in turn, are a few (up to four) orders of magnitude
more luminous than FR\,Is. Following the classic model of Scheuer (1974) the radio source luminosity (P) changes
as some power of the linear size $L$ as the source expands: $P\propto (L)^{-h}$ and Fanti et al. (1995) expect
$h\approx 0.45-0.65$. Assuming that a typical LSO is roughly two orders of magnitude larger than a typical CSS
source and substituting $h=-0.65$ to that formula we get a factor of 20 for the luminosity drop which is
insufficient to make the luminosity evolution of CSS source to FR\,I a likely scenario.

>From the above argument there is hardly any doubt left that strong CSS sources are FR\,II progenitors. Where are
the FR\,I progenitors then? Quite naturally we should look for them among weaker CSS sources but until the end of
the last century the so-called 3CRPW sample (Spencer et al.\ 1989) was the canonical sample of CSS sources used
for most of the research in that field. Only recently a major step towards weaker sources has been made when Fanti
et al.\ (2001) derived a sample of 87 CSS sources from the B3-VLA sample (Vigotti et al.\ 1989) and surveyed them
with the VLA. The most compact objects from that sample have been followed up with VLBI by Dallacasa et al.
(2002a, 2002b). Further work on intermediate luminosity CSS sources from the B2 sample has been done by Saikia et
al. (2003). Another contribution to this area of research comes from our group; we searched the FIRST survey
(White et al.\ 1997) to find many weak CSS sources and followed them up with MERLIN, EVN, and the VLBA (Marecki et
al.\ 2003, hereafter Paper I). Finally, Tschager (2003) used the WENSS catalogue (Rengelink et al.\ 1997) to find
a number of even fainter CSS sources.

Readhead et al.\ (1996) proposed an evolutionary scheme unifying three classes of objects: compact symmetric
objects (CSOs), MSOs, and LSOs. Snellen et al.\ (1999, 2000, 2003) discussed many
aspects of the above scenario in detail. In particular they concluded that the radio luminosity of GPS/CSO objects
increases as they evolve, reaches its maximum for the CSS/MSO phase, and then gradually decreases as these objects
grow further to become LSOs.\footnote{By the way, Snellen et al. (2000) admit it is unclear if {\em all} young
sources evolve to LSOs. We will discuss this issue in the next section.} Snellen et al. (1999, hereafter SSMBRL)
suggest (see their Figure 1) that those LSOs could be either FR\,Is or FR\,IIs. However, we think that the
evidence now suggests that CSS sources evolve {\em only} to FR\,II. Inspection of the maps resulting from all
those recent surveys leads to a conclusion that, again, it is very hard to find a mini-FR\,I among weak CSS
sources. 2358+406 (Dallacasa et al. 2002a) and 1601+382 (Paper I) seem to be exceptions to the rule that CSS
sources are almost exclusively mini-FR\,IIs and there is no clear, direct evolutionary link between CSS sources
and FR\,Is.

So where do FR\,Is come from? This is a timely question and we think the answer to this is just emerging.
Ghisellini \& Celotti (2001) discussed many aspects of the FR\,I/FR\,II dividing line. Since the optical
luminosity of an elliptical galaxy is an indication of the mass of the central supermassive black hole (SMBH) and
since there is a correlation between the luminosities at the FR\,I/FR\,II boundary in the radio and optical
domains, they propose that the FR\,I/FR\,II dividing luminosity is a function of the mass of the SMBH. This
immediately leads to a conclusion that FR\,Is have, on average, larger black hole masses. If so, they might be
older or have accreted at a greater rate in the past and therefore at least a fraction of them have been FR\,IIs
in the past. The above argument is compatible with the finding of an apparent lack of any evolutionary link
between CSS sources and FR\,Is and so the evolutionary sequence of radio-loud AGN might look as follows:

\vspace{0.8cm} \centerline{GPS/CSO $\rightarrow$ CSS/MSO $\rightarrow$ FR\,II $\rightarrow$ FR\,I}

\section{Do All Young Sources Reach the LSO Stage?}

Before Readhead et al. (1996) placed CSOs as an initial stage of evolution of radio-loud AGN they had given
another interpretation of the CSO phenomenon (Readhead et al. 1994). According to that, some or even the majority
of CSOs may be short-lived emitters due to a lack of sufficient fuelling and they `fizzle out' (as the authors
wrote) after a relatively short period of time ($\sim 10^4$\,years). Actually, SSMBRL account for them and label
them `drop-outs' on their Figure 1. That bifurcation in the evolutionary track leads quite naturally to a
suggestion that the evolutionary track outlined above is not necessarily the only one. In fact, a whole family (a
continuum?) of such tracks might exist and the one shown by SSMBRL just appears as the only one simply due to
selection effects.

So the truth could be as follows: the choice of a particular evolutionary track merely depends on the duration of
the activity period of the host galaxy. In the case of LSOs, radio lobes are powered by central engines for the
longest possible time but if the energy supply cuts off much earlier, an object leaves the `main sequence'
proposed by SSMBRL and it will never reach the LSO stage. It stops growing, its luminosity drops, and the spectrum
gradually gets steeper and steeper. Instead of a full-blown LSO we will get a (very) steep spectrum object of
either minute or moderate size.

This conjecture can be firmly underpinned with the theory of the SMBH accretion disc thermal--viscous
instabilities (Siemiginowska, Czerny, \& Kostyunin 1996; Siemiginowska \& Elvis 1997; HSE01). Figures 1 and 2 in
HSE01 explain in a very appealing manner that the mass of the SMBH determines the length of the activity phase of
an AGN as well as the timescale of the activity re-occurrence --- the latter  scales as a square of mass of the
SMBH. In other words it is the SMBH mass that determines the choice of the evolutionary track of an AGN.

Reynolds \& Begelman (1997) independently suggested a model in which extragalactic radio sources are intermittent
on timescales of $\sim10^{4} - 10^{5}$ years. When the power supply cuts off, the radio source fades rapidly in
radio luminosity. However, the shocked matter continues to expand supersonically and keeps the basic source
structure intact. This model predicts that there should be a large number of medium-sized objects (a few hundred
to a few thousand parsecs across) which are roughly by an order of magnitude weaker than MSOs known so far because
of the power cutoff. Modern CSS source surveys (like ours) seem to be sufficiently deep to test this prediction.

So let's consider the following scenario: the energy supply from the central engine cuts off after $\sim 10^5$
years, i.e. a time\-scale typical for spectral ages of CSS sources (Murgia et al. 1999; Murgia 2003). According to
HSE01 it corresponds to an SMBH mass of the order of $10^8 M_{\sun}$. In such circumstances we should get a
`dying' CSS source. How should such an object differ from an ordinary CSS source in the case when both are not
beamed toward us, i.e. we perceive them as MSOs? A dying MSO should be weak because of lack of fuelling, have an
even steeper spectrum because of radiation and expansion losses and, last but not least, it should not be edge
brightened because there are no more (or hardly any) jets pushing through the ISM so the hot spots should (almost)
have faded away. What we should see therefore is nothing more than just diffuse, `amoeba-shaped' lobes without
well-defined hot spots. We think that five sources in our FIRST-based sample of weak CSS sources (1009+408,
1236+327, 1542+323, 1656+391, and 1717+315) are good candidates for dying CSS sources/MSOs (Paper I). As a matter
of fact, LSOs possessing these features --- the so-called `faders'
--- although relatively rare, {\em have} already been observed mainly as a
part of surveying samples with ultra-steep spectrum sources (R\"ottgering et al. 1994; de Breuck et al. 2000).

Moving towards lower masses of the SMBH by one order of magnitude, the length of activity period gets shorter by
two orders of magnitude according to the thermal--viscous instabilities model. This means that the transition to
`coasting' (or `fader') phase would happen at a very early stage of the evolution, i.e. the CSO stage. The large
majority of CSOs exhibit a gigahertz peaked spectrum caused by synchrotron self-absorption in the mini-lobes
located at their extremities and producing the peak at about 1\,GHz. If the fuelling stops prematurely, the hot
spots should fade out and so the spectrum should no longer be featured by that peak. This is a simple way how
CSOs, which usually happen to also be GPS sources, could become non-GPS, ultra-compact CSS sources.

We think we found such ultra-compact steep spectrum objects in our FIRST-based sample. Those sources are {\em not}
GPS sources because of the selection criteria of the sample (see Paper I) but still they are hardly resolved or
completely unresolved by MERLIN at 6\,cm, i.e. they have angular sizes below $\sim 50$\,milliarcseconds. Could
they be the candidates for `dying' CSOs? We have just been allocated global VLBI time to observe them. If we find
symmetric, double VLBI structures with evidence of very weak or non-existent hot spots within the diffuse lobes,
and with no cores (thereby suggesting that the central energy source has turned off), they will automatically
become candidate members of the class being a `dead end' in the evolutionary track of CSOs.

\section{Recycling of Radio Structures of Radio-Loud AGN}

There is yet another obvious consequence of adopting the theory of thermal--viscous instabilities in the accretion
disk. It predicts that the activity phase of an AGN takes $\sim 30\%$ of the time and  is followed by a quiescence
period. The full cycle of life of a galaxy (provided it happens to be radio-loud in its active phase) might
therefore look as follows:

\vspace{0.6cm}

\centerline{ignition of activity $\rightarrow$ GPS/CSO $\rightarrow$ CSS/MSO $\rightarrow$ FR\,II $\rightarrow$
FR\,I} \centerline{$\rightarrow$ decline of activity $\rightarrow$ quiescence.} \vspace{0.6cm}

\noindent For less massive SMBHs that cycle would lack the most advanced
(LSO) stages and so it would shrink to:

\vspace{0.6cm}
\centerline{ignition of activity $\rightarrow$ GPS/CSO $\rightarrow$
CSS/MSO $\rightarrow$ decline of activity $\rightarrow$ quiescence}
\vspace{0.6cm}

\noindent or even to:

\vspace{0.6cm}
\centerline{ignition of activity $\rightarrow$ GPS/CSO $\rightarrow$
decline of activity $\rightarrow$ quiescence.}
\vspace{0.6cm}

Since the quiescent phase takes a significant amount of time (70\% of the activity cycle --- HSE01), it looks that
in principle the ultimate stage of a fader is a total disappearance. However, it is conceivable that the `coasting
phase' could last during the {\em whole} quiescent period or even longer. If so, then it can happen that a new
phase of activity starts {\em before} the lobes have faded completely. The presence of a new, compact double-lobed
radio source located in the centre of a large-scale, also double-lobed relic structure is  observable proof of
such a circumstance. Radio galaxies with the so-called `double--double' morphology have already been detected by
Schoenmakers et al. (1999, 2000). Another spectacular example of the restarted activity of a radio-loud AGN ---
1245+676, the giant radio galaxy (GRG) with a CSO core --- is given by Marecki et al. (2003). Nevertheless,
double--double morphologies are not common just because the quiescent phase of a galaxy takes such a large
fraction of its lifetime and so it is hard for the lobes  in the coasting phase to survive. However, AGN with
relatively low massive SMBHs should recycle at a much faster rate and so the duration of the quiescent phase is
not that long; both double--double and `dying source' morphologies should be more readily observed in these
objects. Moreover, such a fast rate of recycling causing many young sources to decay in an early (e.g. $10^5$
years) or very early (e.g. $10^3$ years) stage of their development and preventing their further growth, could
explain the abundance of compact objects (see e.g. Perucho \& Mart\'{\i} 2003).

\section{Conclusion and Prospects}

There are seemingly many different paths of the evolution of radio-loud AGN. The theory of thermal--viscous
instabilities (HSE01) together with the conclusions drawn by Ghisellini \& Celotti (2001) provide a framework that
explains in a consistent and quite convincing manner why a given AGN follows a particular path and what the
terminal points of those paths could be. In brief: it is all controlled by the mass of the SMBH. For the most
massive SMBHs the object can reach the most advanced stage of the evolution --- an FR\,I. Depending on the SMBH
mass, a decline of activity can take place virtually at any stage of the evolution. Early decline would result in
a `young fader', i.e. a compact object without hot spots within its diffuse lobes. Firm observational confirmation
of young faders and particularly non-GPS, ultra-compact CSS sources would provide a strong underpinning of that
theoretical framework. A possible discovery of compact double--doubles would provide a sound proof that the range
of possible timescales of activity cycle is in fact very large. Those cycles can either be long in the case of
GRGs like, for example, 1245+676, or short for low SMBH mass AGN.

\section*{References}


\reference Akujor, C.E., \& Garrington, S.T.\ 1995, A\&AS, 112, 235

\reference Dallacasa, D., Tinti, S., Fanti, C., Fanti, R., Gregorini, L., Stanghellini, C., \& Vigotti, M.\ 2002a,
A\&A, 389, 115

\reference Dallacasa, D., Fanti, C., Giacintucci, S., Stanghellini, C., Fanti, R., Gregorini, L., \& Vigotti, M.\
2002b, A\&A, 389, 126

\reference de Breuck, C., van Breugel, W., R\"ottgering, H.J.A., \& Miley, G.K.\ 2000, A\&AS, 143, 303

\reference Fanaroff, B.L., \& Riley, J.M.\ 1974, MNRAS, 167, 31P

\reference Fanti, C., Fanti, R., Dallacasa, D., Schilizzi, R.T., Spencer, R.E., \& Stanghellini, C.\ 1995, A\&A,
302, 317

\reference Fanti, C., Pozzi, F., Dallacasa, D., Fanti, R., Gregorini, L., Stanghellini, C., \& Vi\-gotti, M.\
2001, A\&A, 369, 380

\reference Ghisellini, G., \& Celotti, A.\ 2001, A\&A, 379, L1

\reference Hatziminaoglou, E., Siemiginowska, A., \& Elvis, M. 2001, ApJ, 547, 90 (HSE01)


\reference Marecki, A., Niezgoda, J., Wlodarczak, J., Kunert, M., Spencer, R.E., \& Kus, A.J.\ 2003, PASA, 20, in
press (Paper I)

\reference  Murgia, M.\ 2003, PASA, 20, in press

\reference Murgia, M., Fanti, C., Fanti, R., Gregorini, L., Klein, U., Mack, K.-H., \& Vigotti, M.\ 1999, A\&A,
345, 769

\reference Perucho M. \& Mart\'{\i} J.M.\ 2003, PASA, 20, in press

\reference Readhead, A.C.S., Xu, W., Pearson, T.J., Wilkinson, P.N., \& Polatidis, A.G.\ 1994, in Compact
Extragalactic Radio Sources, NRAO Workshop No. 23, ed.\ J.A. Zensus \& K.I. Kellermann

\reference Readhead, A.C.S., Taylor, G.B., Pearson, T.J., \& Wilkinson, P.N.\ 1996, ApJ, 460, 634

\reference Rengelink, R.B., Tang, Y., de Bruyn, A.G., Miley, G.K., Bremer, M.N., R\"ott\-gering, H.J.A., \&
Bremer, M.A.R.\ 1997, A\&AS, 124, 259

\reference Reynolds, C.S., \& Begelman, M.C. 1997, ApJ, 487, L135

\reference R\"ottgering, H.J.A., Lacy, M., Miley, G.K., Chambers, K.C., \& Saunders, R. 1994, A\&AS, 108, 79

\reference Saikia, D.J., Jeyakumar, S., Mantovani, F.,  Salter, C.J., Spencer, R.E., Thomasson,  P., \& Wiita,
P.J.\ 2003, PASA, 20, in press

\reference Scheuer, P.A.G.\ 1974, MNRAS, 166, 513

\reference Schoenmakers, A.P., de Bruyn, A.G., R\"ottgering, H.J.A., \& van der Laan, H.\ 1999, A\&A, 341, 44

\reference Schoenmakers, A.P., de Bruyn, A.G., R\"ottgering, H.J.A., van der Laan, H., \& Kaiser, C.R.\ 2000,
MNRAS, 315, 371

\reference Siemiginowska, A., \& Elvis, M.\ 1997, ApJ, 482, L9

\reference Siemiginowska, A., Czerny, B., \& Kostyunin, V.\ 1996, ApJ, 458, 491

\reference Snellen, I.A.G., Schilizzi, R.T., Miley, G.K., Bremer, M.N., R\"ottgering, H.J.A., \& van Langevelde,
H.J.\ 1999, NewAR, 43, 675 ({\tt astro-ph/9811453}) (SSMBRL)

\reference Snellen, I.A.G., Schilizzi, R.T., Miley, G.K., de Bruyn, A.G., Bremer, M.N., \& R\"ottgering, H.J.A.\
2000, MNRAS, 319, 445

\reference Snellen, I.A.G.,  Mack, K.-H., Schilizzi, R.T., \& Tschager, W.\ 2003, PASA, 20, in press

\reference Spencer, R.E., McDowell, J.C., Charlesworth, M., Fanti, C., Parma, P., \& Peacock, J.A.\ 1989, MNRAS,
240, 657

\reference Tschager, W.\ 2003, PASA, 20, in press

\reference Vigotti, M., Grueff, G., Perley, R., Clark, B.G., \& Bridle, A.H.\ 1989, AJ, 98, 419

\reference White, R.L., Becker, R.H., Helfand, D.J., \& Gregg, M.D.\ 1997, ApJ, 475, 479

\end{document}